\documentclass[fleqn,11pt]{wlscirep}
\usepackage[utf8]{inputenc}
\usepackage[T1]{fontenc}
\usepackage{graphicx}
\usepackage{float} 
\usepackage{setspace}
\usepackage{lineno}

\title{Using Satellite Imagery and Machine Learning to Estimate the Livelihood Impact of Electricity Access}

\author[1]{Nathan Ratledge}
\author[2]{Gabe Cadamuro}
\author[3]{Brandon de la Cuesta}
\author[4]{Matthieu Stigler}
\author[4,5,6]{Marshall Burke}
\affil[1]{Emmett Interdisciplinary Program in Environment and Resources, Stanford University, Palo Alto, CA 94305, USA}
\affil[2]{Atlas AI, Palo Alto, CA 94305, USA}
\affil[3]{King Center for International Development, Stanford University, Palo Alto, CA 94305, USA}
\affil[4]{Center on Food Security and the Environment, Stanford University, Palo Alto, CA 94305, USA}
\affil[5]{Department of Earth System Science, Stanford University, Palo Alto, CA 94305, USA}
\affil[6]{National Bureau of Economic Research, Cambridge MA 02138}


\begin{abstract}
In many regions of the world, sparse data on key economic outcomes inhibits the development, targeting, and evaluation of public policy.  We demonstrate how advancements in satellite imagery and machine learning can help ameliorate these data and inference challenges. In the context of an expansion of the electrical grid across Uganda, we show how a combination of satellite imagery and computer vision can be used to develop local-level livelihood measurements appropriate for inferring the causal impact of electricity access on livelihoods. We then show how ML-based inference techniques deliver more reliable estimates of the causal impact of electrification than traditional alternatives when applied to these data. We estimate that grid access improves village-level asset wealth in rural Uganda by 0.17 standard deviations, more than doubling the growth rate over our study period relative to untreated areas.  Our results provide country-scale evidence on the impact of a key infrastructure investment, and provide a low-cost, generalizable approach to future policy evaluation in data sparse environments. 

\end{abstract}
\begin{document}

\flushbottom
\maketitle

\thispagestyle{empty}

\onehalfspacing

\section*{Introduction}
\indent Accurate public policy evaluation requires data on outcomes of interest that are likely to be shaped by a given policy, and an ability to isolate the impact of the policy from other correlated changes that might also affect these outcomes.  Sparse data on key livelihood outcomes across much of the world continues to hinder such evaluations \cite{Burke:2021}. In many countries, existing data provide only an intermittent or aggregate snapshot of well-being, and the near absence of granular and reliable livelihood data over time makes it challenging to both measure changes in well-being and to correctly attribute them to policy interventions. Left unaddressed, this data shortage will continue to impede understanding of what interventions best improve livelihoods in some of the most impoverished regions of the world.  

\indent The combination of recent advancements in computer vision and a growing abundance of satellite imagery have recently shown promise in improving the measurement of livelihoods in data-sparse environments\cite{Burke:2021, Yeh:2020, Jean:2016, chi2021micro,Steele:2017}. Deep learning models trained on daytime satellite imagery and nighttime lights have been shown capable of predicting up to 75\% of the variation in local-level economic data on which the model was not trained\cite{Yeh:2020, Steele:2017, Pokhriyal:2017}. But, with few exceptions (eg ref\cite{Huang:2020}), these new approaches have yet to be deployed in the evaluation of policies or interventions designed to improve livelihoods\cite{Burke:2021}.

\indent Successfully deploying such approaches to measure policy impact involves surmounting at least four challenges.  First, as most policy interventions are not randomized, reliable measurement of policy impact typically requires observation of both treated and untreated units (e.g. individuals or locations) before and after the policy was instituted. Generating such longitudinal data from a remote sensing and deep learning pipeline, particularly at a scale and resolution necessary to evaluate a policy's causal effect, is challenging and has not been the focus of previous work\cite{Burke:2021}. A second concern is that image-derived proxies for a dependent variable of interest could themselves include the policy intervention of interest, undermining inference.  For example, poverty predictions from a satellite-based model cannot reliably be used to study the impact of new road construction on poverty if there is a chance that the model looks for a road to decide whether a location is poor. Third, for continuous outcomes, the distribution of predicted outcomes from both simple linear models and more complex deep learning models tend to have lower variance than the distribution of observed outcomes. Models seeking to predict economic well-being, for example, often overpredict for poorer individuals and underpredict for wealthier ones. If interventions are targeted at certain parts of the outcome distribution (e.g. at the poorest) or have heterogeneous effects across it, then this bias in outcome measurement could bias estimates of treatment effects, as we show herein.  Finally, and relatedly, recipients of interventions often have outcomes that are evolving differentially relative to untargeted populations -- for instance, road improvement projects could be targeted to areas with rapid economic growth.  As a result, benchmark research designs, such as difference-in-differences (DD), will likely struggle to recover reliable treatment effect estimates when outcomes -- even when reliably measured -- trend differentially across treated and untreated groups in the pre-treatment period \cite{Jaeger:2018, Kahn-Lang:2019}. 

\begin{figure}[H]
    \centering
    \includegraphics[width=\linewidth]{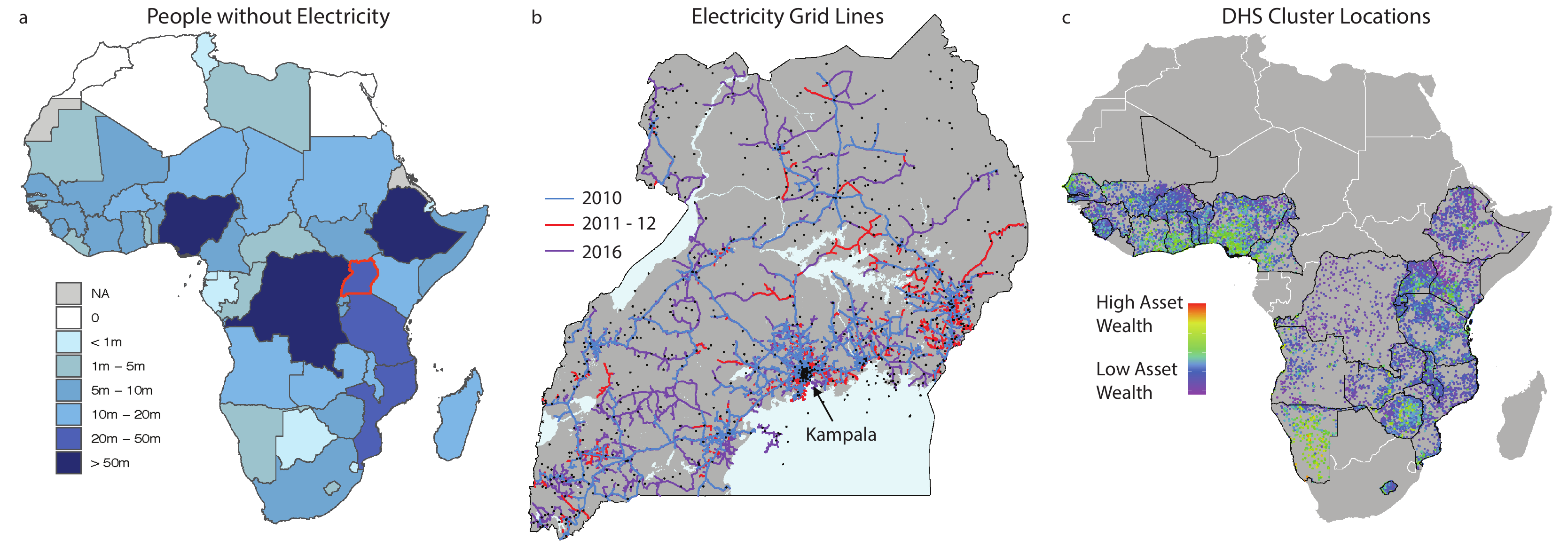}
    \caption{\textbf{Electricity access rates, grid extensions in Uganda and DHS survey locations}. \textbf{a}, Nearly 600m people in Sub Saharan Africa lacked access to modern electricity in 2018. Countries are colored by the number of people lacking access to central grid-based power. Uganda, with roughly 24m without power in 2018, is highlighted by a red border. \textbf{b}, The electricity grid has expanded rapidly in Uganda in recent years, including to new regions of the country. Black dots illustrate the 641 enumeration areas from the 2016 DHS survey. \textbf{c}, Wealth indexes for 27,174 DHS enumeration areas are colored from low asset wealth to high asset wealth. Our training data covers 25 countries over 14 years and represents 641,621 household surveys.} 
    \label{fig:Fig. 1}
\end{figure}

\indent Here, we develop methods to overcome these multiple data and inference challenges and apply them to estimate the causal impact of rural electrification expansion on livelihoods at the country scale in Uganda.  Nearly 1b people globally and 600m in Sub Saharan Africa (SSA) lack access to modern electricity (Fig. \ref{fig:Fig. 1}a) \cite{WEO:2019, AfricaEO:2019}, and the roughly ~\$20b invested annually in expanding electricity access in SSA in recent years is only a fraction of the estimated \$45-100b needed annually over the next 20 years to secure continent-wide access\cite{AfricaEO:2019}. The likely impact of this past expansion is debated, and the benefits of future expansion uncertain. While some research finds positive effects of electrification on livelihoods\cite{Dinkelman:2011}, recent experimental work finds limited impacts\cite{Lee:2020_b, Lee:2015}, and other studies note the lack of consistent evidence\cite{Bayer:2020}. Understanding where and under what conditions expanded electricity access is most beneficial will be critical for guiding the proposed ~\$1-2.5T investment in SSA electrification efforts in coming decades\cite{AfricaEO:2019}. 

\indent To study the effect of electrification on livelihoods in Uganda, we combine georeferenced data on the multi-year expansion of the electricity distribution grid throughout the country (Fig. \ref{fig:Fig. 1}b) with satellite-based predictions of asset wealth from a machine learning (ML) model trained on data from 27k villages across SSA (Fig. \ref{fig:Fig. 1}c). We then use two ML-based econometric approaches to isolate the average treatment effect (ATE) of grid expansion on asset wealth in the largely rural and peri-urban areas that received grid access during our study period. 

\indent Specifically, we construct data on grid expansion by combining publicly available data from the Uganda government, the World Bank and partners, and academic papers\cite{Omulo:2015, Uganda:2015, Uganda:2016} (Methods). Our harmonized data on grid locations covers the years 2005, 06, 10, 13-18. To estimate livelihood impacts, we focus on locations that received grid access in 2011 or 2012, a period of rapid grid expansion where we can observe outcomes for many years both before and after expansion. We define ``treated'' communities as those that were within 2km of new distribution lines in 2011 or 2012. Control communities are defined as those that did not receive grid access by 2016, the most recent year for which Ugandan DHS survey data is available (Methods).  

\indent We measure outcomes using data on household-level asset wealth, a reliable and commonly used indicator of economic well-being\cite{filmer2001estimating} that is consistently measured in many georeferenced nationally-representative surveys conducted in SSA.  Unfortunately, such surveys -- in particular the Demographic and Health Surveys (DHS) -- do not revisit the same households or locations across survey waves, making it difficult to construct the repeated local-level measurements on which many causal inference techniques depend. However, recent studies have shown that convolutional neural networks (CNNs) can be trained to accurately predict asset wealth at the village or neighborhood level from satellite imagery \cite{Yeh:2020,chi2021micro}, and that such predictions can be used to fill in data gaps in both space and time. This gap filling can enable subsequent application of causal inference approaches that require longitudinal data.

\indent Building on this work, we assemble survey-based household asset wealth data from DHS surveys of 641,621 households in 27,174 enumeration areas (EA) across 25 countries in SSA over 14 years (Fig. \ref{fig:Fig. 1}c, Extended Data Fig. \ref{fig:DHSsurveys}).  Following prior studies\cite{filmer2001estimating,Yeh:2020}, we use principal components analysis (PCA) to construct an asset wealth index (WI) from household's survey responses to questions about ownership of specific assets (Methods, Extended Data Fig. \ref{fig:DHSvariables}), using data from all households in the sample to construct one common index. We then create a mean WI for each ``cluster" or EA (akin to a town or village) by averaging household-level WIs within each cluster.  We focus on asset wealth rather than other livelihood measurements (e.g. consumption expenditure) as asset wealth is considered a less-noisy measure of households’ longer-run economic well-being, is a common component of multi-dimensional livelihood measures used around the world\cite{sahn2003exploring, filmer2012assessing}, and is well measured in our survey data. 

\indent We then train a CNN to predict wealth index values at the cluster level from temporally and spatially matched multispectral satellite images. We use a ResNet-18 architecture\cite{He:2016} with the input layer modified to include additional satellite bands to supplement RGB imagery, and the final layer modified to provide a single scalar estimate of the wealth composite. We split data into disjoint training (60\%), validation (20\%), and test sets (20\%), with the model evaluated on the held-out test data (Methods). We evaluate CNN models trained with and without data from Uganda in the training and validation set; the latter setting replicates a common real-world situation in which ground-truth data from a target geography of interest may not exist.

\indent To overcome the concern that visual indicators of the independent variable of interest (electricity distribution grid) could be used by the CNN in its construction of the wealth index and thus generate a mechanical relationship between grid expansion and asset wealth, we take two steps. First, departing from past work, we exclude measures of self-reported household electrification from the construction of the WI; this had only minor effects on WI estimates (Extended Data Fig. \ref{fig:WI_comparison}). Second, we used only daytime satellite imagery rather than nighttime lights imagery as input to the CNN, as nighttime imagery directly can pick up the presence of local electrification, and the coarse (30m) daytime imagery that we use is unlikely to directly detect electrification infrastructure.

In our setting, the use of standard mean squared error loss functions in CNNs (or in simpler linear models) generates a distribution of model-predicted values that is lower variance than the true distribution (Extended Data Fig. \ref{fig:Bias_correction}a), with the model over-predicting wealth at the lower end of the wealth distribution and underpredicting in wealthier regions.  We show mathematically and through simulation that this bias is consequential for downstream inference tasks, attenuating (biasing towards zero) treatment effect estimates that use these predictions to measure outcomes (Methods, Extended Data Fig. \ref{fig:DD_MC_plot}, c).  To mitigate these biases, we modify the standard MSE loss function with an additional term that penalizes bias in each quintile of the wealth distribution (Methods). This loss function generates wealth predictions that do not lead to attenuated estimates in downstream inference tasks (Extended Data Fig. \ref{fig:Bias_correction}), but at the cost of somewhat lower predictive performance as measured by standard test statistics (e.g. $r^2$, a common test statistic of interest in this setting\cite{Yeh:2020,Burke:2021}).  This result highlights that the common practice of maximizing average predictive performance across the output distribution (e.g. for a regression problem, maximizing the $r^2$ between predicted and observed values in a test set) could actually worsen performance on key downstream tasks. 

Our custom loss function helps ensure that bias in outcome prediction will not lead to bias in estimated treatment effects. But accurate outcome data alone do not solve the separate, more conventional inference problem in which a given unit is only observed in either a treated or untreated state, and the unobserved counterfactual state must be inferred to estimate the effect of treatment. To solve this inference task, we exploit two recent ML-based causal inference approaches, matrix completion (MC) and synthetic controls with elastic net (SC-EN) which have been shown to be more robust in the face of common threats to inference such as non-parallel trends in pre-treatment outcomes\cite{Athey:2021, Doudchenko:2016}.  MC works by treating counterfactual untreated observations in the treatment group as missing values in a matrix, with these values imputed through a regularized process that penalizes matrix complexity\cite{Athey:2021} (Methods). MC's main identifying assumption is that predicted values (in the matrix of outcomes values) are uncorrelated with treatment\cite{Athey:2021, Xu:2017}.  SC-EN uses a flexible approach to matching treated units with a weighted average of control units who were trending similarly pre-treatment\cite{Doudchenko:2016}. For causal identification, SC requires that only treated units are affected by the intervention and only post-treatment, and that a weighted combination of untreated units can approximate a treated unit's outcome absent treatment\cite{Abadie:2021}.  We use simulation and cross-validation to show that MC and SC-EN are less biased in the presence of time-trending unobservables and capably predict held-out observations in control locations or pre-treatment observations in treated areas (Methods, (Extended Data Fig. \ref{fig:DD_MC_plot},a, Extended Data Fig. \ref{fig:Cross_validation})). 

\indent We then estimate causal impacts in three separate scenarios. In our base case ("with-Uganda"), DHS data from Uganda contribute to the training and validation data used to train the CNN, mimicking a setting where some local data are available to train models. In our "without-Uganda" case,  we train on DHS data from African countries other than Uganda, mimicking a setting where no local data are available for model development in the target geography.  Predictions in these locations are made for all study years 2006-16. In the third ("full country") case, instead of filling in missing years for the available DHS locations we use the CNN to predict wealth at 1km over all Uganda and all study years  (Fig. \ref{fig:Fig. 3}), masking out unpopulated areas using data on inhabited locations\cite{NASA:2020}. Of these $\sim$6,900 village locations, 3,235 are unelectrified by 2016 and are used as controls and 209 receive electrification during the 2011-2012 period and are used as treated units, resulting in a total of 3,444 observations in the full country analysis. This third case replicates a likely deployment scenario in which a model is trained on any available data and then used to predict values for all locations and years that are unobserved. 

\section*{Results}
\subsection*{CNN Predictive Performance}
Our CNN model's performance across SSA consistently achieved $r^2$ values above 0.6 with a max of 0.69, thus explaining roughly 2/3rds of the variation in held-out village-level DHS survey measurements. Performance on this wealth prediction task is comparable to recent literature, which reports $r^2$ values between 0.6 and 0.75 and typically uses additional inputs (e.g. nighttime lights) that we did not use\cite{Yeh:2020, Burke:2021}. Our preferred CNN model with our bias correction penalty had a combined $r^2$ of 0.63 over all of SSA and a 0.63 $r^2$ for Uganda clusters. In the without-Uganda setting we found slightly lower $r^2$ values when predicting on held-out Ugandan data, with  an $r^2$ of 0.61 across SSA and 0.50 within Uganda. 

\indent "Full-country" predictions are shown in Fig. \ref{fig:Fig. 2}a,b, and changes in wealth during the study period for populated areas are shown in Fig. \ref{fig:Fig. 2}c. Wealth estimates indicate the North and East of the country remain the poorest parts of the country, consistent with World Bank analyses\cite{WB_UBOS:2018}. Our model predictions also indicate substantial wealth growth in central Uganda, in particular the peri-urban areas near the capital city of Kampala, and slower growth in northern regions with a recent legacy of conflict, consistent with recent studies\cite{WB:2015}. On average, our village WI's increased by 0.19 units between 2006 and 2016, equal to a 0.15 sd increase relative to the pooled Ugandan wealth distribution between 2006 and 2016.

\subsection*{Estimated impacts of electrification}
\indent MC and SC-EN estimate similar, statistically significant effects of electrification on wealth across our three data imputation settings (Fig. \ref{fig:Fig. 3}).  In the with-Uganda setting MC estimates a 0.22 [95\% CI 0.08, 0.37] unit improvement in asset-based wealth for treated communities, equal to a 0.17 sd increase relative to the observed Ugandan wealth distribution. Results are robust to alternate inclusion buffers (Extended Data Fig. \ref{fig:3km_4km_results}), and as expected point estimates are attenuated when CNN prediction bias at different parts of the wealth distribution is less heavily penalized, producing a larger Berkson-type error (Extended Data Fig. \ref{fig:Comparitive_treatment_effects}). Treatment effects appear to increase over time (Fig. \ref{fig:Fig. 3}b), consistent with a commonly held hypothesis that it may take time for electricity's impact to be realized \cite{Dinkelman:2011, burlig2016out}.  

\indent Estimates from a model that did not use Uganda data in training are nearly identical to the with-Uganda case, and using wealth predictions across all of Uganda yields a modestly lower but much more precisely estimated effect [0.19 , 95\% CI 0.11,0.26].  The substantially narrower confidence interval in this latter setting reflects the larger amount of data available when all populated locations and years are used to estimate treatment effects.  Both results suggest that treatment effects can be accurately estimated even absent training or validation data in the target geography.

\indent We compare our main results with two DD estimates that could have been implemented absent our satellite-derived data: a `two-unit' DD where outcomes for each treatment group and period are constructed as the simple average of wealth values across any treated or control unit observed in that period's survey (recall that the same locations are not observed across survey rounds), or an ``inverse-distance-weighted" (IDW) DD that imputes wealth values in missing location-years as the relevant inverse-distance-weighted average of nearby treated or control surveyed locations in that year.  Given the moderate failure of DD on the pre-treatment placebo test described above, in neither case is unbiasedness guaranteed. In the two-unit case, the treatment effect point estimate is similar to our main estimate but with a wide confidence interval that contains zero. In the IDW case, the treatment effect estimate is similar in both magnitude and statistical significance to our main estimate.  

\indent We also compare our main results to output from a CNN model with a typical mean squared error loss function (i.e. without the additional bias penalty); recall this prediction model had a higher overall prediction $r^2$. Using predictions from this model, we estimate a 0.13 [95\% CI -0.04,0.24] effect of electrification for the with Uganda case (Extended Data Fig. \ref{fig:Comparitive_treatment_effects}), or roughly 50\% smaller than our preferred estimate and not statistically significant at conventional levels. This result again highlights the importance of optimizing the prediction model for the downstream task.

\section*{Discussion}  
Across multiple approaches to generating wealth estimates (with-Uganda, without-Uganda, full country) and multiple machine learning-based estimators, our results suggest that 
villages who gain electricity grid access enjoy increases in asset wealth as compared to non-electrified communities. Our with-Uganda estimates suggest that treated communities grew more than twice as fast during the post-treatment period compared to unconnected communities. Our broader, full country data setting suggests similar and more precisely estimated impacts on asset wealth. These accelerated growth rates for communities with new access to electricity are particularly notable in the communities that we study, which at baseline are some of the least wealthy on the African continent. Our results suggest that, at least in our Ugandan setting, access to electrification can generate meaningful improvements in economic wellbeing. 

Our findings stand somewhat in contrast to recent experimental work in nearby Kenya that found limited impact of household-level electrification on livelihoods\cite{Lee:2020_a,Lee:2020_b}.  One possible explanation for this difference is that we study, in effect, the household impact of community-level electrification, whereas the Kenya work examined the impact of individual households gaining grid access. Community-level electrification could unlock a range of economic benefits and positive local labor-market spillovers that might not be observed in a household-level intervention\cite{Lee:2020_b}. Our findings, with some exceptions\cite{burlig2016out}, are more in keeping with other quasi-experimental work that found positive effects of community-level grid expansion, including studies in South Africa\cite{Dinkelman:2011}, the Philippines\cite{chakravorty2016lighting}, and Brazil\cite{lipscomb2013development}.

\indent Our results do not provide clear insight on the many potential mechanisms through which expansion of the electricity grid could improve household asset wealth. Indeed, our univariate prediction framework is not well positioned on its own to provide such insight.  Such insight could potentially come by combining our comprehensive outcome data with information on key mediating variables (e.g. labor force participation, agricultural incomes, etc).  

\indent More broadly, our findings illustrate that careful combination of satellite imagery and deep learning can generate data appropriate for use in downstream causal inference tasks, including program evaluation. We believe our two-stage machine learning strategy -- combining imagery and deep learning to generate outcome measurements, and applying new ML-based causal inference estimators to these data to estimate program treatment effects -- represents a scalable, generalizable and relatively low-cost strategy for expanding policy evaluation in data sparse environments, especially where such evaluation was previously cost-prohibitive or infeasible. We show that our strategy performs well even in a challenging setting in which no data from the target country of interest are available for model training. Estimates from our approach can be updated with relative ease as new satellite imagery and grid data become available. 

\indent Our multi-tiered methodology could likely be further improved by additional ground truth surveys and/or additional inputs, such as higher resolution or more frequent satellite imagery. However, we believe the current availability of satellite imagery and state of deep learning methods is more than adequate for making progress on a series of challenging questions related to the causes or consequences of infrastructure development, agriculture productivity improvements, public health interventions, and broad-scale environmental change, and that their combination represents a new frontier for policy evaluation. Finally, our results point to the importance of carefully considering ``downstream" causal inference tasks when selecting an ``upstream" model that will feed data to those tasks. In the setting considered here, customary approaches to optimizing prediction performance can substantially worsen bias in the causal estimate of interest, a dynamic we expect exists in many related applications. 

\paragraph{Acknowledgements}
We thank seminar participants at Stanford and AtlasAI for helpful comments, and colleagues in Uganda for their help in locating and verifying the electricity grid maps. N.R thanks the TomKat Center for Sustainable Energy at Stanford for financial support. 

\paragraph{Author contributions statement}
N.R. conceived of the study, collected the data, led the econometric analysis and wrote the paper. G.C. designed and performed the CNN modeling, developed the bias penalty term and wrote the paper. B.Q. contributed to econometric analysis and wrote the paper. M.S. contributed to the econometric analysis and wrote the paper. M.B. advised the project, contributed to econometric analysis and wrote the paper.  

\paragraph{Competing interests}
M.B. is a cofounder at Atlas AI, a company that uses machine learning to measure economic outcomes in the developing world. G.C. is an employee at Atlas AI. 

\bibliography{MLbibtex}

\newpage

\newpage

\begin{figure}[H]
    \centering
    \includegraphics[width=\linewidth]{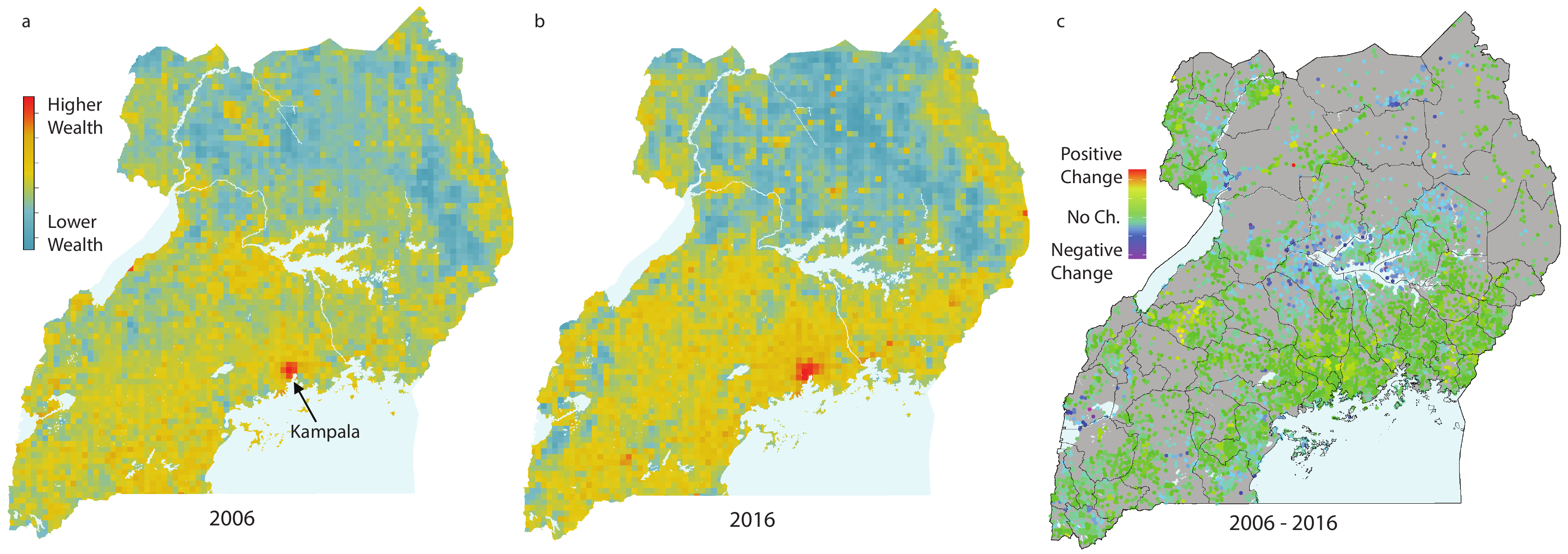}
    \caption{\textbf{CNN-based wealth predictions over time in Uganda}. \textbf{a, b}, Asset-based wealth index values predicted via CNN across Uganda in 2006 and 2016, the first and last years of our analysis. \textbf{c}, Shows the change in asset wealth between 2006 and 2016 for 6,900 villages and neighborhoods. Across all observed locations, asset-based wealth increased by an average of  11\% over the period.}
    \label{fig:Fig. 2}
\end{figure}

\newpage

\begin{figure}[H]
    \centering
    \includegraphics[width=\linewidth]{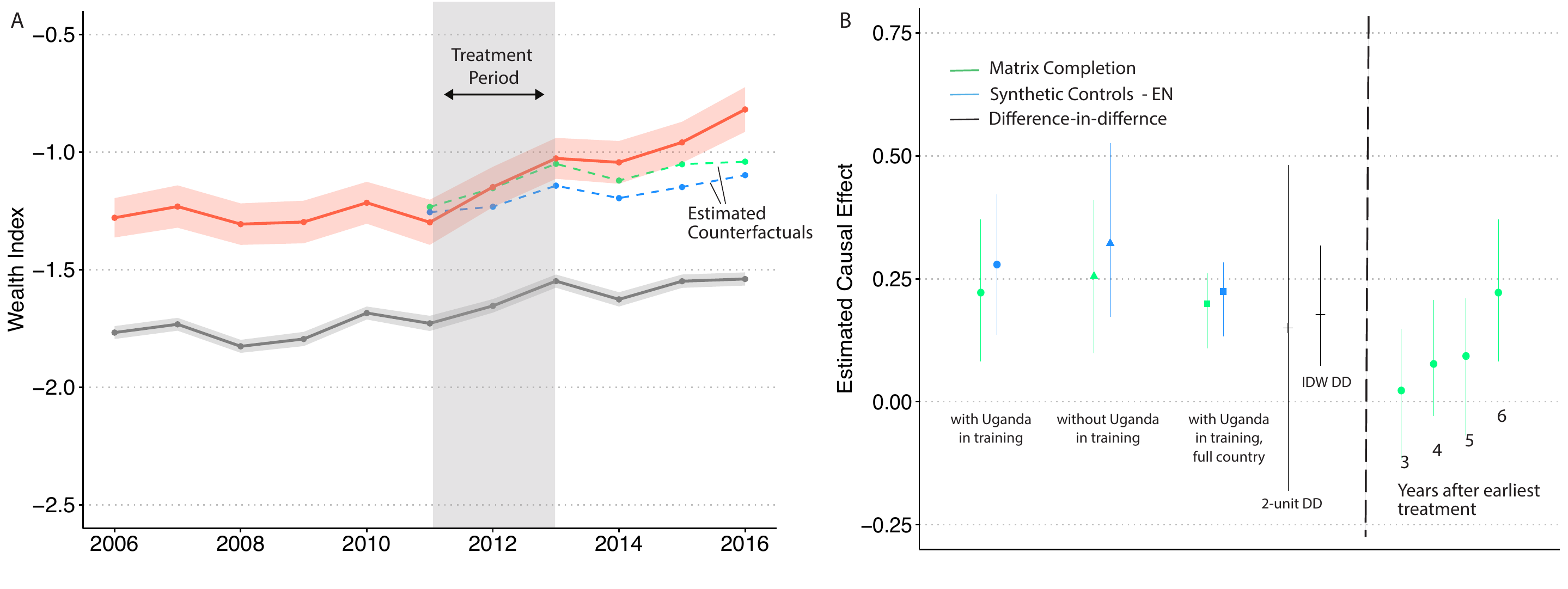}
    \caption{\textbf{Electricity access increases household wealth when compared to households in unelectrified communities}.  \textbf{a}, Average asset wealth in control (gray) and treated (red) locations, bands represent the standard error of the mean. Counterfactuals predicted by matrix completion and synthetic controls with elastic net are shown by dotted lines. \textbf{b}, Estimated causal effect of electrification on wealth at the end of the sample (2016). Error bars represent 95\% confidence intervals. Confidence intervals for each of ML estimates are based on 100 bootstrapped model runs. We find similar statistically significant positive impacts in each of our three ML-based estimates (with Uganda, without Uganda, and full-country). '2-unit' represents a repeated cross-section difference-in-difference (DD) run using only DHS survey estimates. 'IDW' reflects DD results from a inverse distance weighting approach. The four numbered lines at right represent the causal estimates in the 3rd - 6th years (2013 - 2016) after the initial treatment year.} 
    \label{fig:Fig. 3}
\end{figure}

\newpage

\section*{Methods}

\section{Dataset construction}
\subsection{Creating an asset-based wealth index}
\indent The Demographic and Health Surveys (DHS) are a set of nationally-representative household surveys conducted periodically in many low and middle-income countries around the world.  Among other things, the DHS ask a relatively consistent set of questions of household asset ownership and related housing characteristics, including questions on water supply, ownership of assets like a car, and floor, roof and wall materials. Variables such as floor type are converted from descriptions of the asset to a 1–5 score indicating the quality of the asset. We then construct an asset wealth index (WI) at the household level from the first principal component of survey responses to a consistent set of 13 questions (Extended Data Fig. \ref{fig:DHSvariables}). Such an index is a standard approach in development economics to measuring longer-term well-being\cite{sahn2003exploring,filmer2001estimating} and is meant to capture household asset ownership as a single dimension, rather than act as a direct measure of consumption poverty. By construction, the WI has a mean equal to 0 and standard deviation of 1 across all households. 

\indent Included DHS data come from 25 countries in Sub Saharan Africa (SSA) between 2005 and 2017, totalling 641,621 households in 27,174 enumeration areas (EAs, roughly villages in rural areas or neighborhoods in urban areas) (Extended Data Fig. \ref{fig:DHSsurveys}). In Uganda, we use five DHS surveys (2006, 2009, 2011, 2014, 2016), amounting to 1,798 EAs. EAs are geolocated in DHS surveys; however, each reported location includes random jitter (spatial offset) to protect anonymity. Urban clusters include up to 2km displacement, and rural areas commonly up to 5km and sometimes 10km, which introduces random noise into our data setting but does not substantially effect predictive quality\cite{YehSupp:2020}. Before using the DHS data in model training or causal inference, we average household wealth indexes at the EA level.

\indent Because of our focus on measuring the impact of electricity access, we do not include DHS's binary ``has electricity" (HV206) survey question in our PCA.  We withhold this variable to avoid mechanically embedding our independent variable of interest in our dependent variable. However, because ownership of individual household assets and housing characteristics tend to be highly correlated, excluding this input variable does not meaningfully change our WI. We compare our base WI to an index that includes the electricity variable, an index that has the binary electricity variable but not the electrical appliances and an index that has neither the electricity variable nor electrical appliances (Extended Data Fig. \ref{fig:WI_comparison}). We find high correlation between the WI we use and these alternative indexes, with $r^2$s of .99, .97 and .96 respectively.

\subsection{Constructing the electric grid time series}
To create the electricity grid time series, we collected data from a variety of publicly available sources, including published government documents, academic studies, and aid agency reports that mapped Ugandan's electrical distribution system. In total we collect grid data for the years 2005, 2006, 2010, and 2013 to 2018, and explicitly use data from 2010, 2013 and 2016\cite{Omulo:2015, Uganda:2015, Uganda:2016}. To verify accuracy of these maps we conducted in person interviews in Kampala with utility professionals and aid agencies and contacted report authors.  

\indent Data from 2016 and later was made available in digital format from the Ugandan government, in collaboration with the World Bank and GIZ.  Older grid data is extracted from non-digitized maps via map rectification. We accomplish this task by uploading static images to ArcGIS, extracting grid lines and converting them to a digital latitude and longitude format.  To further confirm the accuracy of the earlier, non-digitized maps, we develop a `back casting' strategy where, under the assumption that once the grid is built it rarely disappears, we map the most recent years first and then use these maps to inform grid locations in earlier years.  For example, with the 2016 grid mapped, we could overlay an earlier year and have the primary aim of tracking were the grid stopped in that year, rather than mapping that entire earlier year by itself.  To our knowledge our resulting grid expansion data set is the first national-level time series of an electric grid's evolution in SSA. 

\subsection{Identifying treatment and control groups}
We denote communities (either DHS EA locations or villages from across Uganda) as "treated" if they received distribution grid access between 2011 and 2012. Our treatment does not claim that all homes or businesses received and used grid-based electricity. Rather, our treatment assumes the community received grid access. To identify our treatment and control groups, we apply a 2km buffer to our electricity grid lines. (We test results when using a 3km or 4km buffer (Extended Data Fig. \ref{fig:3km_4km_results}) and find similar treatment effects). Using the time series of grid development we create a list of EAs (or villages in the full country setting) that were connected pre-2011, during 2011 and 2012, or post 2012.  Locations that received grid access between 2011 and 2012 comprise our treatment group.  We also identify sites that were not connected by 2016. These never-connected villages are our control group.

\indent In the full country setting where we apply the CNN across Uganda rather than just in DHS locations, we modestly update our treatment selection methodology. First, because there is no random jitter in village coordinates, we shrink our treatment range to 1km on either side of the grid line. Second, as many villages are tightly clustered in this data setting, we add a 1km buffer outside of the 1km treatment range, meaning that all control units are still beyond 2km from the grid line. Finally, we add a population density layer to our village map\cite{OSM:2019}, and drop any village where the population density layer is equal to zero (n = 4).  We also drop the highest 1\% of villages by population density from our treatment group (n = 31), as these are above three standard deviations from the mean and we consider these observations to be potential outliers. In this setting, we end up with 209 treated villages and 3,235 control villages, each being predicted annually from 2006 to 2016. 

\section{Wealth prediction}

\subsection{Satellite-based wealth estimation in the context of downstream inference tasks}

We wish to infer the impact of electrification on livelihoods, using data on livelihoods predicted by a model. For continuous outcomes, the distribution of predicted outcomes from both simple linear models and more complex deep learning models tend to have lower variance than the distribution of observed outcomes. This is mechanically true for linear regression models, and is empirically true for deep learning models in our setting (Extended Data Fig. \ref{fig:Bias_correction}a). In practice, this means that predicted wealth is overstated for poorer segments of the population and understated for wealthier segments of the population, meaning prediction bias is a function of underlying wealth.   

We explore whether this so-called Berkson error can be consequential for downstream causal inference tasks in our setting. Consider a standard two period difference-in-difference (DD) estimation, with true wealth values $y$ measured in both periods for treatment and control units, and predicted (potentially biased) wealth values $y'$ in the same units and periods. Assume that predicted wealth is a linear transformation of true wealth, i.e. $y' = \alpha + \phi y$, where in general we expect $\alpha > 0$ and $0 < \phi < 1$.  Predicted wealth is too high at low wealth levels and too low at high wealth levels. 

For notational simplicity, we label period and group-specific true and observed wealth values as (e.g.) $T_1$ and $T'_1$ for true and predicted wealth in the treatment group in the treated period, and do not take expectations\footnote{We use this simplified notation for the sake of clarity. Strictly speaking,  $T_1$ refers to the expected value of treated units in the post-treatment period observed without error, i.e. $T_1= E[y|D=1, T=2]$. Likewise, $T'_1$ is the same quantity yet observed with measurement error: $T'_1= E[y'|D=1, T=2]$.}.  The "true" DD estimate is given by:
\begin{equation}
    \beta = (T_1 - T_0) - (C_1 - C_0)
\end{equation}
The DD estimated on predicted data is:
\begin{equation}
    \beta' = (T_1' - T'_0) - (C'_1 - C'_0)
\end{equation}
Using the mapping from true to predicted wealth, we can then write the DD bias as:
\begin{align}
    \beta' - \beta & = [(T'_1 - T'_0) - (C'_1-C'_0)] -  [(T_1 - T_0) - (C_1 - C_0)] \nonumber\\
    & = [((\alpha + \phi T'_1) - (\alpha + \phi T'_0)) - ((\alpha + \phi C'_1) - (\alpha + \phi C'_0))] -  [(T_1 - T_0) - (C_1 - C_0)]\nonumber \\
    & = [\phi(T_1 - T_0) - \phi(C_1 - C_0)] - [(T_1 - T_0) - (C_1 - C_0)] \nonumber\\
    & = (\phi - 1)\beta
\end{align}
Bias is only a function of the slope parameter $\phi$ and not intercept $\alpha$ (any level bias nets out in DD).  Because we expect $0 < \phi < 1$, then our DD estimate will be attenuated:  downward biased if true effect is positive, and upward biased if true effect is negative in our two period setting. We thus seek model predictions where $\phi$ is nearest to 1 as possible, i.e. where Berkson error is eliminated.

\subsection{Wealth estimation with a convolutional neural net.}
We seek a prediction model that can accurately predict local-level wealth using inputs that are available comprehensively across space and time, and who's predictions do not suffer from the Berkson-type error described above. Coarse-resolution public satellite imagery from Landsat and other sensors is one such available input, and prior work has shown that a convolutional neural network (CNN) can be trained to accurately predict local-level asset wealth using Landsat as input\cite{Yeh:2020,Burke:2021}.  However, the use of standard CNN loss functions that minimize MSE across the entire output distribution tend to generate wealth predictions with the Berkson-type error described above. 

We thus develop and embed a custom loss function in a ResNet-18 CNN architecture that is able to generate wealth predictions from satellite inputs without Berkson error. Our basic approach is to additionally penalize bias at separate quintiles of wealth distribution.  
This reduced bias comes at the expense of overall predictive performance (e.g. as measured by $r^2$), and, generally speaking, the best models on $r^2$ grounds tend to also produce the largest Berkson error. This suggests that researchers wishing to use data generated from CNN and other machine learning architectures should tailor their loss function according to their desired downstream inference task.  

We use a ResNet-18 architecture\cite{He:2016} with a modified input layer to accommodate our multi-spectral input data (224x224x6) and a modified final layer to produce a single continuous estimate instead of predictions for multiple classes. The model is trained from scratch using an Adam optimizer\cite{Kingma:2017} with a learning rate decay of 0.96 per epoch. We split our data into 5 equally sized subsets and set up five different random folds in which 3 of the subsets (60\%) are assigned to the training set and one subset (20\%) is assigned to the validation and test sets each. The validation data is used to tune the model while final performance statistics are calculated using the test set. We ensure that all five subsets are used as the validation or test set exactly once.

\subsection{Custom loss function}
Define the 5 quintiles of the survey-derived wealth distribution as $Q_1$ to $Q_5$, with $Q_1$ being the lowest quintile of wealth. Consider a mini-batch (a subset of the data used in each training step) with $K$ training data points $X_i$ with wealth labels $y_i$. The CNN at this current step can be defined as $\hat{f}$ and the predicted wealth given by $\hat{y} = \hat{f}(X_i)$. We seek to compute the bias $B$ of the estimator $\hat{f}$ with respect to each quintile of wealth, and then penalize some aggregation of this bias. Ideally we would measure $B_{1}(\hat{f})$ ... $B_{5}(\hat{f})$ as the bias that $(\hat{f})$ generates over all the data points in $Q_1 \ldots Q_5$. 

However, in practice we are constrained to using the data in each minibatch, and instead compute the \textit{sample bias}  $\hat{B}_{1}(\hat{f}), \ldots, \hat{B}_{5}(\hat{f})$ for each quintile as follows:
\begin{equation}
    \hat{B}_{j}(\hat{f}) = \mathbb{E}[\hat{f}(X_i) - y_i | y_i \in Q_j]
\end{equation}
Using these quintile-specific bias estimates, we then define our custom loss function: 
\begin{equation} \label{eqn:custom_loss_empirical}
    \hat{E}_b = \max_{j}(\hat{B}_{j}(\hat{f})^{2})
\end{equation}
with $\hat{B}_{j}(\hat{f})$ the estimate of the true bias of the CNN on the $j^{th}$ quintile $B_{j}(\hat{f})$. This estimation of the bias will hence have an error associated with it that grows smaller as the number of samples fed to it (the $y_i$ and $\hat{f}(X_i)$ ) increases. This is why we increase the batch size to 90, to allow for an average of 18 samples per quintile. Equation \ref{eqn:custom_loss_empirical} forces the model to minimize the maximum quintile-specific bias.

We then combine this quintile-specific loss function with the standard MSE loss and regularization loss. We write the total loss equation for a given minibatch as:
\begin{equation}
    \mathcal{L} = MSE + \lambda_r L_2 + \lambda_b\hat{E}_b
\end{equation}
where $L_2$ represents the L2 norm of the CNN weights (a standard practice to avoid overfitting). Finding the optimal value of  $\lambda_b$ requires a different approach since it seeks to lower a specific type of bias at the cost of overall MSE, so any search based simply on the standard MSE metric would have returned $\lambda_b =0$. 
Instead, we choose $\lambda_b$ empirically to yield a $\phi$ (the slope of the regression of true on predicted wealth) as close to 1 as possible. Extended Data Figure \ref{fig:Bias_correction} demonstrates the impact of using no quintile-specific bias ($\lambda_b =0$) versus increasingly weighting the quintile-specific bias term ($\lambda_b$ up to 7.5). Empirically we find $\lambda_b = 5$ to yield predictions with $\phi$ closest to 1 (Extended Data Fig. \ref{fig:Bias_correction}d), and so we choose this penalty as our primary model.  We again note that this reduced bias comes at the cost of an overall reduction in prediction $r^2$ (Extended Data Fig. \ref{fig:Bias_correction}f), relative to a model with a standard loss function ($\lambda_b =0$).

Our final CNN parameters are an initial learning rate of .0001, a regularization term of .0001, and a bias penalty term of 5.  Using these model parameters we produce five disjoint splits that estimate a WI for every DHS location or grid cell in Uganda on an annual basis from 2006 to 2016. With five splits, this means that every year-unit observation will have five independent estimates. 

\subsection{Imagery inputs}
Earlier efforts to predict local-level well-being from imagery typically have used a combination of daytime and nighttime imagery\cite{Yeh:2020,Jean:2016}. In our analysis, we use only daytime imagery from Landsat as input, as we are concerned that nighttime light imagery was likely a direct proxy for electrification and thus its inclusion (much like the inclusion of household electrification in the asset index) could mechanically embed our independent variable of interested into our dependent variable of interest. 

We utilize 6 bands in Landsat: the standard three RGB visual bands, one near-infrared band and two shortwave infrared bands. All of the bands have a native resolution of 30 meters per pixel and so our standard CNN input patch size of 224 pixels means that each input image approximately covered a 6.72x6.72 square kilometer area. Our goal is to obtain data for each pixel in each year. As LANDSAT has many images at a given location for a year we generated a composite by taking the median pixel value over all non-cloudy images falling within the year span. Taking a composite in such a way enables us to operate on a single image (as opposed to the more complex case of a sequence of images) and conceptually removes confounding seasonal weather effects or other scene-specific artifacts. Compositing was done in Google Earth Engine. 

\subsection{Training with and without Uganda}
While our target geography is Uganda, we have wealth data from dozens of other African countries on which to train our CNN. Past studies have suggested models trained in one African country tend to perform well in nearby countries\cite{Yeh:2020,Jean:2016}, and that -- for daytime-imagery-based models in particular -- having many training observations improves model performance\cite{Yeh:2020}. 

We thus train our CNN using data from countries outside Uganda. 
In our core ``with-Uganda" scenario, we also allow observed DHS clusters from Uganda to be randomly selected into the CNN training and validation sets, which intuitively could increase performance in our target geography of interest.  We also create a ``without Uganda" setting where we do not allow the CNN to see any Uganda data in the training and validation sets.  This is designed to replicate a challenging but common real world situation in which target geographies may have no data for training or validation. Our preferred bias-corrected CNN models in the with-Uganda case has an average $r^2$ of 0.63 across SSA, the without-Uganda case is 0.61.

In our ``full country" setting we use the same CNN parameters as our base with-Uganda case (initial learning rate of .0001, a regularization term of .0001, and a bias penalty term of 5) and allow the CNN to train on Ugandan DHS clusters.  We then impute WIs across \textit{all} of Uganda from 2006 to 2016 (Fig. \ref{fig:Fig. 2}a,b). Next, we utilize a geolocated dataframe of Ugandan villages and neighborhoods\cite{OSM:2019} and extract wealth indexes for each village on an annual basis. We ultimately end up with $\sim$6,900 village estimates per year (Fig. \ref{fig:Fig. 2}c).

\section{Causal inference}
We explore the use of three different causal inference techniques, the well-known difference-in-differences (DD) estimator, and two machine learning (ML) approaches - matrix completion (MC) and synthetic controls with elastic net (SC-EN)\cite{Athey:2021, Doudchenko:2016}. We evaluate their appropriateness in various data settings. For example, at one end of the spectrum we test a two unit (treatment and control), repeated cross-section (t = 5) case based solely on observed DHS data (n = 1,798) (Fig. \ref{fig:Fig. 3}). At the other end of the spectrum we explore test appropriateness across eleven years with ~3,444 village estimates per year (Fig. \ref{fig:Fig. 2}, Fig. \ref{fig:Fig. 3}). 

\subsubsection*{Difference in difference}
DD estimates the impact of a treatment by subtracting changes in outcomes in an untreated group from changes in outcomes from a newly-treated group. Most DD estimators are used in a panel setting in which both control and (eventually) treated groups are repeatedly observed. In the DHS survey data, the sampling frame changes between survey rounds, meaning that the same villages are not observed in separate surveys. However, absent panel data, DD could still be implemented in two ways on these survey data, and provide a relevant benchmark against which to understand other estimators and results. First, we can simply compute the DD estimator over the observed villages in each round, defining in each period the control villages that were never treated and the treated villages that were eventually treated, and computing mean outcomes over each group. We then stack the control and treated averages across periods and implement DD as a two-way fixed effects regression: 

\begin{equation}
Y_{it} = \beta \tau_{it} + u_i + v_t + \epsilon_{it}
\end{equation}

\noindent where \emph{$Y_{it}$} is the cluster level WI, \emph{$\tau$} is a binary indicator for whether the grid was present, \emph{u} are unit fixed effects, \emph{v} are time fixed effects, and \emph{$\epsilon$} captures the remaining error.  Reported estimates of $\beta$ are shown as "2-unit DD" in Fig. \ref{fig:Fig. 3}b.

A second approach is to interpolate survey observations and use interpolated values to create a village-level panel. To implement this approach, we take all 1,798 clusters locations that were observed in any of the 5 DHS surveys in Uganda (2006, 2009, 2011, 2014, 2016), and in survey years in which that location was unobserved, we impute the wealth value in that location-year as the inverse-distance-weighted average of wealth values in nearby villages, using all observations within 10km of the location in the average; locations without observations within 10km are dropped. We then run our standard two-way fixed effect DD regression and report bootstrapped standard errors ("IDW DD" in Fig. \ref{fig:Fig. 3}b). We note that our DD setting does not have staggered adoption and only has never-treated units as controls, and thus avoids the issues associated with DD highlighted by a number of recent papers\cite{GoodmanBacon:2021,CallawaySantAnna2020}.

\subsubsection*{Testing parallel trends}
Causal identification in DD requires a parallel trends assumption, i.e. that outcomes in control and treatment groups would have trended similarly absent treatment\cite{Jaeger:2018, Kahn-Lang:2019}. A standard check on this assumption is to evaluate whether outcomes were evolving similarly prior to treatment. In these tests we use panel data imputed by our CNN model, not observed DHS survey data. We conduct each test by re-labeling treated units in 2010, our last pre-treatment year, as being treated in that year and run the above DD regression. If the DD estimator is statistically different from zero (p-value < .05 at 95\% CI), we reject parallel pre-trends. We conduct this parallel trends test on a series of different iterations of our data, including randomized CNN split values, as we use in the final causal analysis. The pre-trends tests failed slightly more than would be expected by chance in the setting where Uganda data were used in training, and substantially more than would be expected where Uganda data were not used in training  (Extended Data Fig. \ref{fig:Pretrend_table}).

\subsubsection*{Synthetic controls with elastic net and matrix completion}
\indent Synthetic Controls with Elastic Net (SC-EN)\cite{Doudchenko:2016} is a regularized form of the synthetic control method\cite{Abadie:2010} which uses elastic net (a combination of LASSO and ridge penalties) to limit overfitting when matching treated units to control units. In general, SC-EN is a more flexible model than traditional synthetic controls because it allows weights to be negative and does not require them to sum to zero: Formally, SC-EN estimates the unit specific counterfactual as:

   \begin{equation}
\hat{Y}_{j,T}(0) = \hat{\mu}^{en}(j;\alpha,\lambda) + \sum\hat{\omega}^{en}_i (j;\alpha,\lambda) * Y^{obs}_{i,T}.
   \end{equation}

\noindent $\hat{Y}$ is the counterfactual being predicted. $\hat{\mu}$ is an intercept. $\alpha$ is how much weight to put on the LASSO or ridge components of the elastic net estimator, which we fix at .5, $\lambda$ weights the penalty function and is determined via cross validation, where we find the tuning parameter, $\lambda$, that minimizes the error in cross validation tests\cite{Doudchenko:2016}, and $\omega$ is a weight for each control observation, $Y^{obs}_{i,T}$. In words, SC-EN predicts the counterfactual, $Y_{cf,post}$, by weighting each post-period control variable with $\omega_i$, noting that zero weights are common.  These weights are determined via panel like regression within the pre-treatment period, where we are regressing a single treated unit on the full panel of control units. In this setting the control weights are updated for each treated unit. 

\indent In our setting we have fewer time periods (T) than we do units/villages (N), so we run a transposed or perpendicular version of SC-EN following ref\cite{Athey:2021}. Thus, we are regressing individual post treatment control years on the panel of pre-treatment controls and estimating year weights, rather than unit weights. We apply these year weights to each pre-treatment treated-unit time series to estimate each unit's year specific counterfactual. 

\indent As in SC-EN, matrix completion\cite{Athey:2021} attempts to impute (unobserved) untreated outcome values for treated units, but uses a matrix decomposition approach for imputation.  

\begin{equation}
       \hat{\mathbf{L}} = arg min_{L}\Bigg[\sum\limits_{(i,t)\in\vartheta} \frac{(Y_{it}-L_{it})^2}{|\vartheta|} +  \lambda||\mathbf{L}||_*\Bigg]
   \end{equation}

\noindent $Y_{it}$ is a matrix (i.e. panel) of observed values (N x T), in our case local wealth indexes in Uganda. It includes both control and treated units pre and post-treatment.  Our goal is to find $\hat{L}$, a regularized estimate of the outcome matrix $Y$.  We want to estimate a penalized version of $Y$ to avoid overfitting, and ultimately predict counterfactuals.  $\lambda||\textbf{L}||$ is the penalty term. As an example, consider a square N x T matrix where only the bottom right cell is `treated'. First, we set our single treated  cell in $Y$ to 0, keeping all the other values as they were observed. Then, MC decomposes this matrix via singular value decomposition, where $Y_{N x T} = S_{N x N} * \Sigma_{N x T} * R^t_{T x T}$, and where $\Sigma$ is the diagonal of singular values. These singular values are penalized by $\lambda$ in a similar fashion as a regularized regression, which forces the singular values closer to zero. Each new, penalized matrix, $L_{it}$, is reconstructed by multiplying $S * \Sigma_{penalized} * R$. The optimal $\hat{L}$ is identified by the minimal root mean squared error between $Y_{it}$ and $L_{it}$. We then use the estimated bottom right cell of $\hat{L}$ as our counterfactual untreated observation for the treated unit. 

Whereas DD requires that treated and control units be trending similarly absent treatment, and is thus biased in the presence of time-trending unobservables, both MC and SC-EN only require that some subset of control units be trending similarly to a given treated unit. These newer estimators are thus in principle more robust to pre-trends violations, and have been demonstrated to perform well in a range of prediction settings\cite{Athey:2021} (Extended Data Fig. \ref{fig:DD_MC_plot}). 

\subsubsection*{Estimator validation}
We perform a series of tests on SC-EN and MC to evaluate their performance in our setting. First, we use $k$-fold cross validation to evaluate the precision of the SC-EN and MC models and their performance against DD in predicting held-out post-treatment values in the control set. We accomplish this by splitting the Uganda control villages into ten equally sized partitions, repeatedly holding out one fold, setting post-2010 values to missing for villages in this fold, and using the remaining 90\% of control observations to impute these missing values using SC-EN, MC or DD for prediction.  Every fold is used as a test set one time and in the training set $k-1$ times, resulting in a unique yearly estimate for each control unit in the post-treatment period.  We then compare each observed control unit value $(y_{it}^c)$ to each predicted control unit $(\hat{y}_{it}^c)$ and compute the RMSE. We also compare the annual observed mean to the annual imputed mean, which reflects average model accuracy, our primary goal. We find that across all post-treatment years the mean difference is 0.002 for DD, MC and SC-EN. RMSE's are 0.36, 0.45, and 0.44, respectively. Cross-validation results are shown in Extended Data Fig. \ref{fig:Cross_validation}. 

Second, and relatedly, we assess how well DD, MC and SC-EN predict outcomes for treated units in 2010, the last pre-treatment year. If outcomes were trending differentially between treated and control groups prior to treatment, then an ability to accurately predict outcomes in the control group (our first test above) would not guarantee an ability to predict counterfactual outcomes in the treated group. To evaluate this latter ability, we randomly select control and treatment units, with replacement, into a sample data set.  To assess DD, we run a standard pre-trends test and record the mean difference from zero (i.e. the DD coefficient). To assess MC and SC-EN, we predict outcome values for treated units in 2010 and compare predicted to observed.  We run this analysis 100 times on a random selection of control and treated units and average the results. We find that MC performs similarly to DD in terms of average prediction error (-0.035 to -0.033) and SC-EN performs modestly better (0.027), despite already limited set of pre-treatment years upon which it predicts. 

Finally, we explore the impact of parallel trends violations on our different estimators.  Non-parallel pre-trends are a plausible concern in our data -- for example, the concern that electrification is targeted to locations that were already growing most quickly -- and as shown in above DD tests is perhaps a practical concern as well. We explore a simulated multi-year setting where treatment has a positive effect (equal to 1) but where treated units pre-treatment are on average trending positively relative to control units (i.e. parallel trends is violated). We then estimate treatment effects using DD and MC, varying the number of years in the sample  (t = 4-20), where treatment can occur at the earliest in (t/2+1)) varying the magnitude of pre-trends violation. We find that MC outperforms DD in most scenarios (Extended Data Fig. \ref{fig:DD_MC_plot}), particularly as pre-trends violations increase or the sample exceeds a few years. 

Overall, we find MC to be better estimator than DD in two of our robustness checks and does similarly well in cross validation, where DD should perform well. Recognizing the other literature highlighting MC's performance\cite{Athey:2021}, we select MC as our preferred causal estimator. We also report SC-EN in our results because it performs quite well in the most common cross validation test. 

\subsection*{Estimating the average treatment effect}
To calculate average treatment effects and their uncertainty, we use a common bootstrap procedure where we randomly select control and treated units with replacement into a sample data set, before executing our causal models (MC, SC-EN, or DD). However, to capture imputation uncertainty, every bootstrap run starts by randomly selecting a single unit-year WI for each of our control and treatment units (recall that our CNN procedure splits our training data into five folds and thus produces five unit-year estimates for each location). Using our base 2km setup as an example, we then randomly select 888 control and 76 treated units to create a run-specific sample data set. Next, we execute our MC and SC-EN procedures, recording each estimated counterfactual for treated units from 2011 to 2016.  As MC re-estimates the entire sample data set, we get all of the estimated counterfactuals in one matrix. For SC-EN we run our algorithm separately for each post-treatment year.  Then, to estimate the year-unit specific treatment effect we subtract the estimated counterfactual from the observed value. To estimate annual ATEs, we average the within-year estimates.  We repeat this analysis on 100 distinct data sets and report bootstrap standard errors at the 95\% level of confidence for both MC and SC-EN (Fig. \ref{fig:Fig. 3}). 

Our core results - for both the with and without Uganda cases - are based on a 2km buffer around the grid line and use a penalty term of 5.  We also report 3km and 4km buffer results using the same penalty term of 5 and find similar results (Extended Data Fig. \ref{fig:3km_4km_results}). And, we estimate results when the penalty term is set at 0, 1, 3 and 7.5 (Extended Data Fig. \ref{fig:Comparitive_treatment_effects}). We find similar effects in all of these cases, consistent with our main results. 

\subsection*{Research related carbon emissions}
\indent We recognize deep learning requires large amounts of computational power and energy, typically resulting in CO\textsubscript{2}e emissions from energy use. At the same time, ML can save emissions if it offsets more consumptive data aggregation strategies. While the artificial intelligence community should be keenly aware of its consumption and emissions, we argue that in our setting there is likely a net benefit to generating asset data using ML algorithms, as machine learning-based predictions offset in person data solicitation, as is done in DHS surveys. For example, the Ugandan DHS surveys we use include 1,798 EAs across five years, whereas our imputed data fills in gaps to produce a net of 19,778 EA across 11 years, more than 10 times as much data.  Thus, assuming some reasonable level of additional in-person-based surveys would be required to conduct a similar analysis (without using ML), it is likely that our ML approach will produce lower emissions.  In our setting, as the heaviest computational work was done on the cloud via Google's data centers, our total emissions profile benefits from Google's progressive clean energy commitments, as well as Stanford University's.  Increasing clean energy penetration will further this advantage and reduce the ML community's ecological footprint.

\subsection*{Data Availability.}
Data and R code to replicate our results figures can be found on GitHub at https://github.com/nwrat?tab=repositories.  

\newpage
\section*{Extended Data Figures}
\setcounter{figure}{0} 

\begin{figure}[H]
\renewcommand{\figurename}{Extended Data Figure}
    \centering
    \includegraphics[width=16cm]{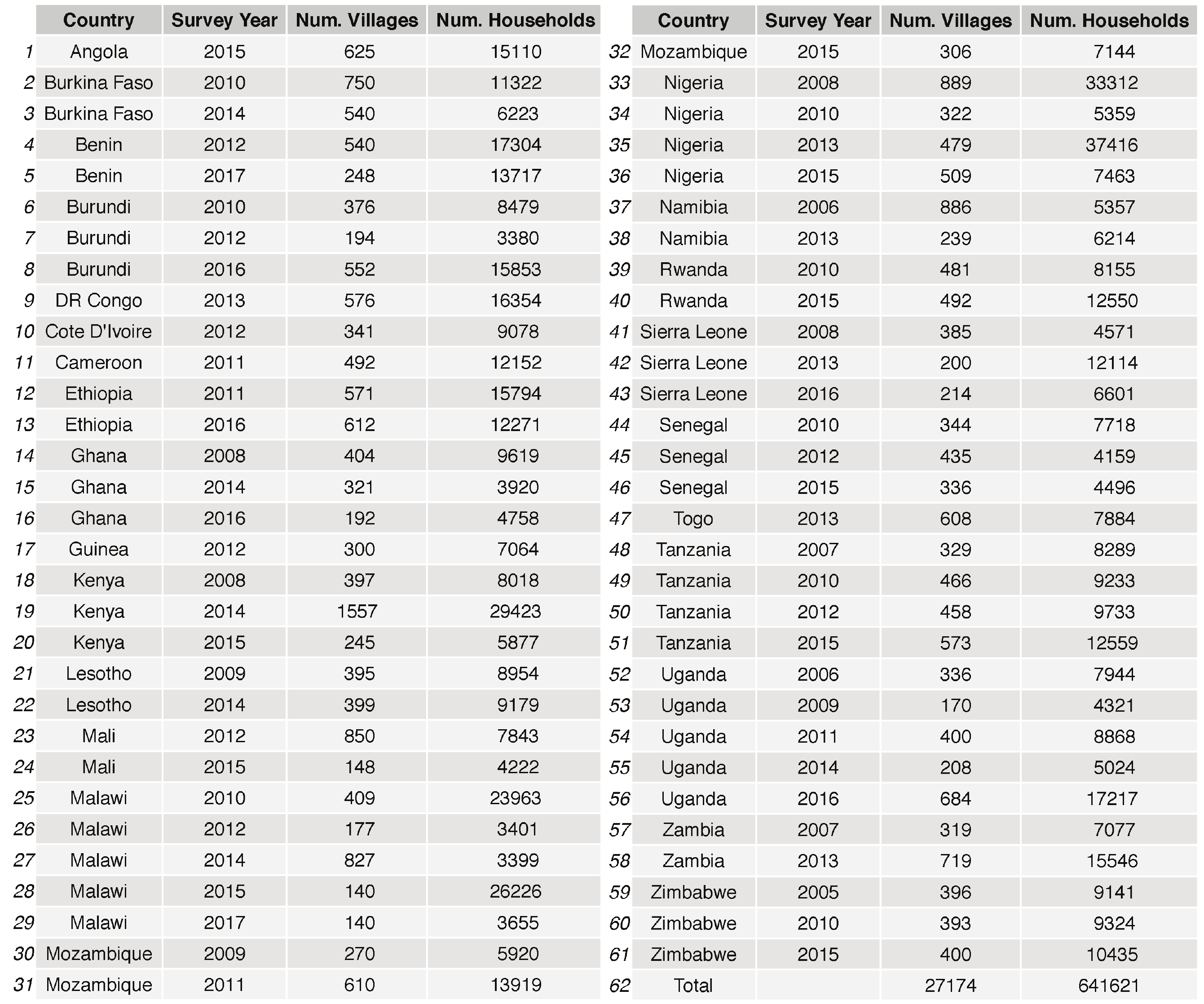}
    \caption{\textbf{61 DHS surveys from Sub Saharan Africa are used in this study}. DHS survey years run from 2005 - 2017. We use asset-based wealth estimates from 27,174 villages, representing data from over 640k households, to train our CNN models.}
    \label{fig:DHSsurveys}
\end{figure}

\begin{figure}[H]
\renewcommand{\figurename}{Extended Data Figure}
    \centering
    \includegraphics[width=6cm]{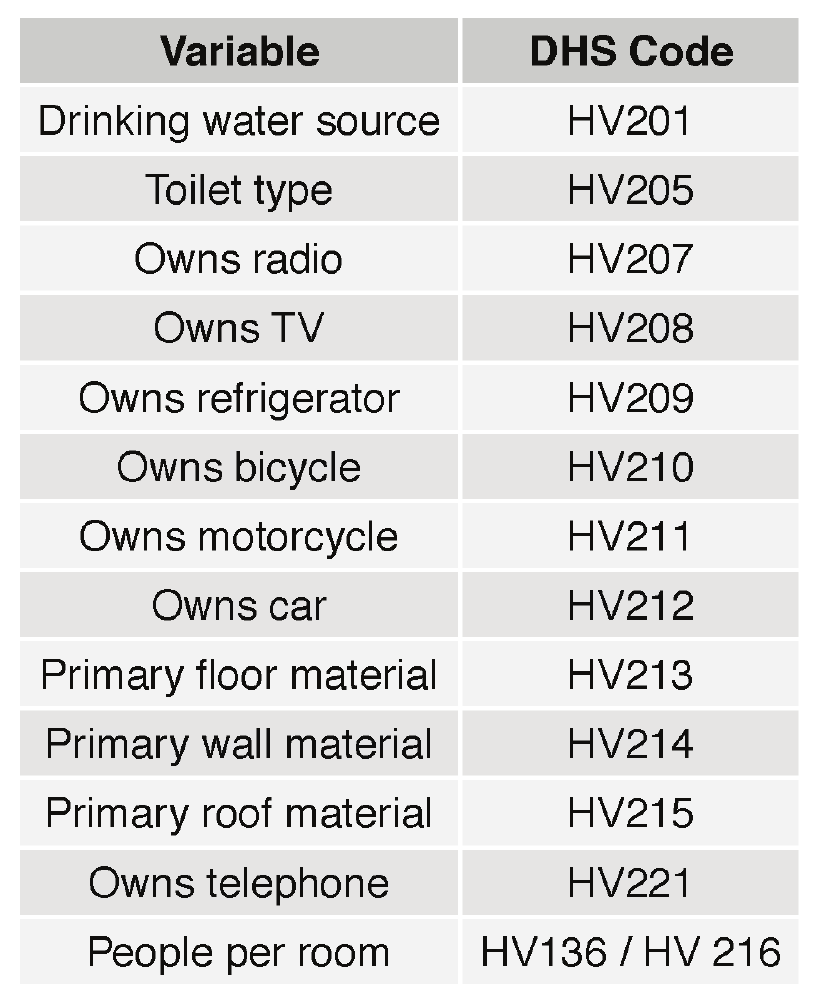}
    \caption{\textbf{DHS variables used in creation of asset wealth index}. }
    \label{fig:DHSvariables}
\end{figure}

\begin{figure}[H]
\renewcommand{\figurename}{Extended Data Figure}
    \centering
    \includegraphics[width=\linewidth]{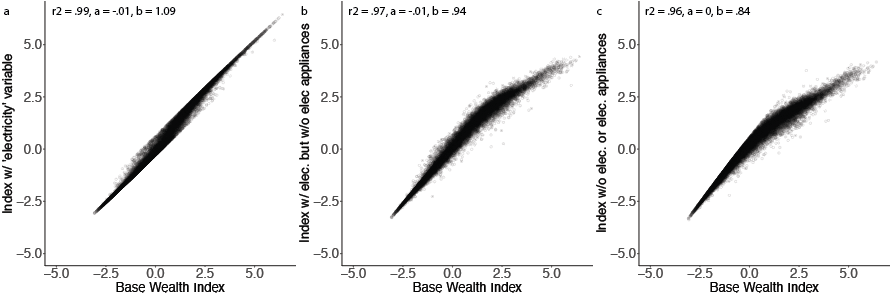}
    \caption{\textbf{Alternate approaches to constructing the village-level wealth index generate highly correlated indices.} \textbf{a}, Compares the base wealth index constructed from the variables listed in ED Fig \ref{fig:DHSvariables} to an index that additionally includes DHS's `has electricity' variable. The $r^2$, intercept and coefficient from regressing the alternative index on the base index are shown. \textbf{b}, Compares the base index to the base index plus the electricity variable but minus three electrical appliances (TV, refrigerator, and phone). \textbf{c}, Compares the base index to the base index minus three electrical appliances.}
    \label{fig:WI_comparison}
\end{figure}

\begin{figure}[H]
    \centering
    \renewcommand{\figurename}{Extended Data Figure}
    \includegraphics[width=\linewidth]{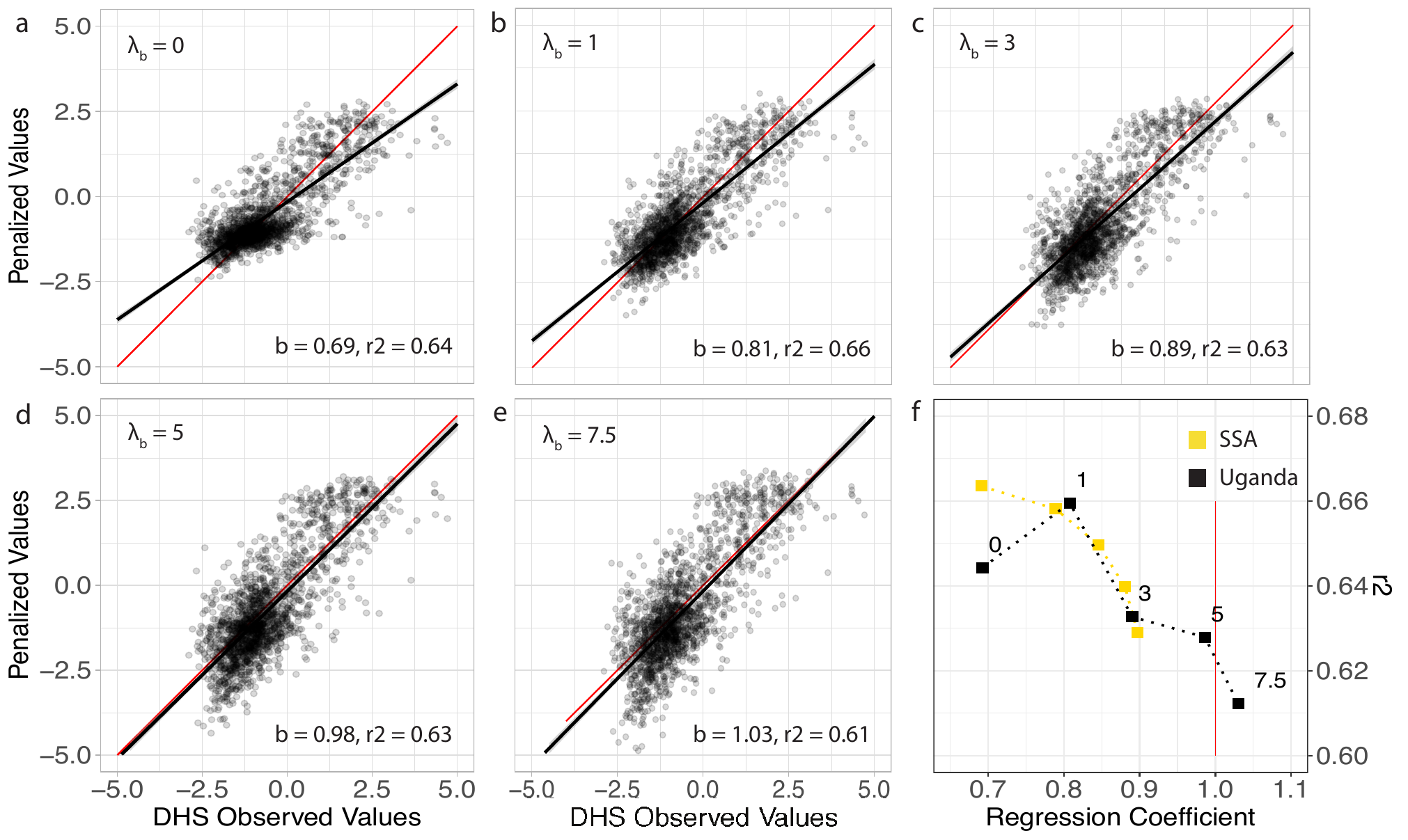}
    \caption{\textbf{Our custom loss function corrects for prediction biases at different points in the wealth distribution, but modestly reduces overall predictive performance ($r^2$) in held-out Uganda data}. \textbf{a-e} DHS observed wealth index values in Uganda on the x axis (n = 1,798) and CNN-predicted values on the y axis. The red line is a 45 degree line, the black is the line of best fit from regressing predicted on observed; slope coefficient and regression $r^2$ are shown in the lower right corner of each subplot.  Higher penalties on quintile-specific bias leads to slope coefficients closer to 1 and slight reductions in $r^2$. However, when the penalty term gets too large, as in the 7.5 case, we see the regression coefficient begin to deviate from 1 again. \textbf{f}, Relationship between the slope coefficient and $r^2$ for each subplot. The results from the entire Sub Saharan data set are shown in gold.} 
    \label{fig:Bias_correction}
\end{figure}

\begin{figure}[H]
\renewcommand{\figurename}{Extended Data Figure}
    \centering
    \includegraphics[width=12cm]{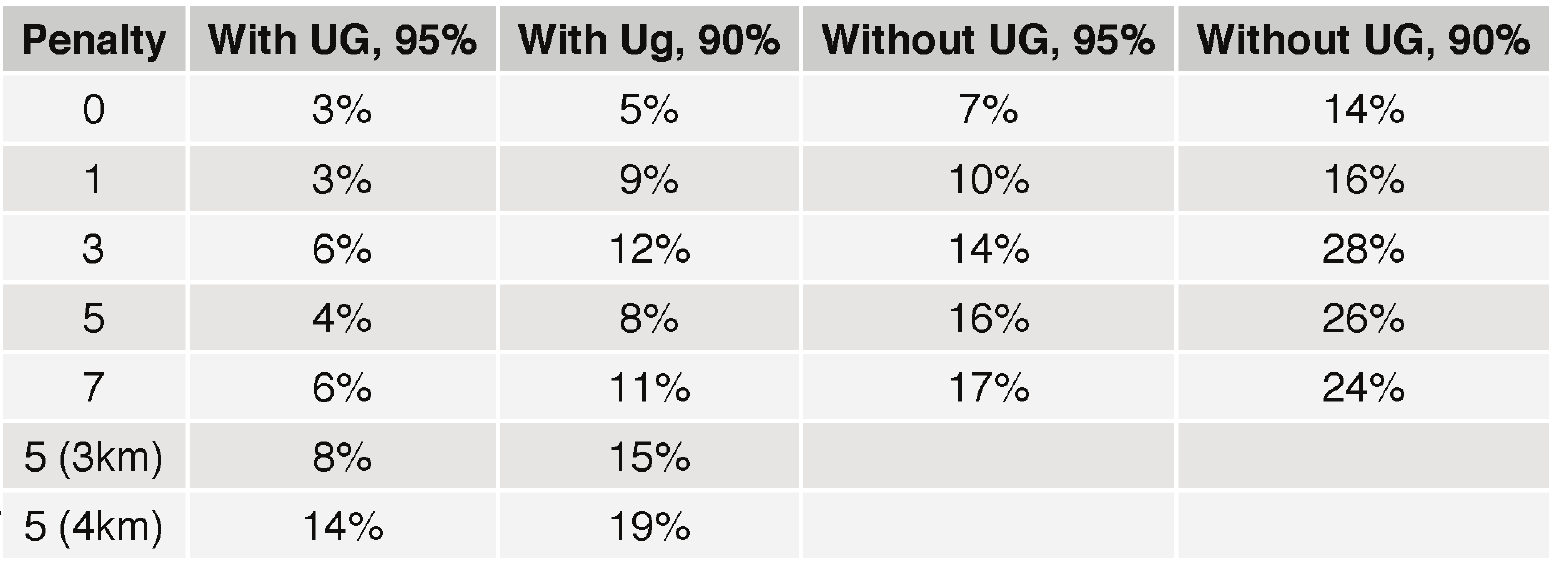}
    \caption{\textbf{Rejection rate for difference-in-difference estimator in pre-trends test}. Percent of runs in which a difference-in-difference estimator rejected a null hypothesis of no pre-trends at either 95\% or 90\% confidence. "With Uganda" and "Without Uganda" reference whether or not Uganda was used in the training and validation steps. "Penalty" indicates the quintile-bias penalty term ($\lambda_b$) used in each model run.}
    \label{fig:Pretrend_table}
\end{figure}

\begin{figure}[H]
\renewcommand{\figurename}{Extended Data Figure}
    \centering
    \includegraphics[width=\linewidth]{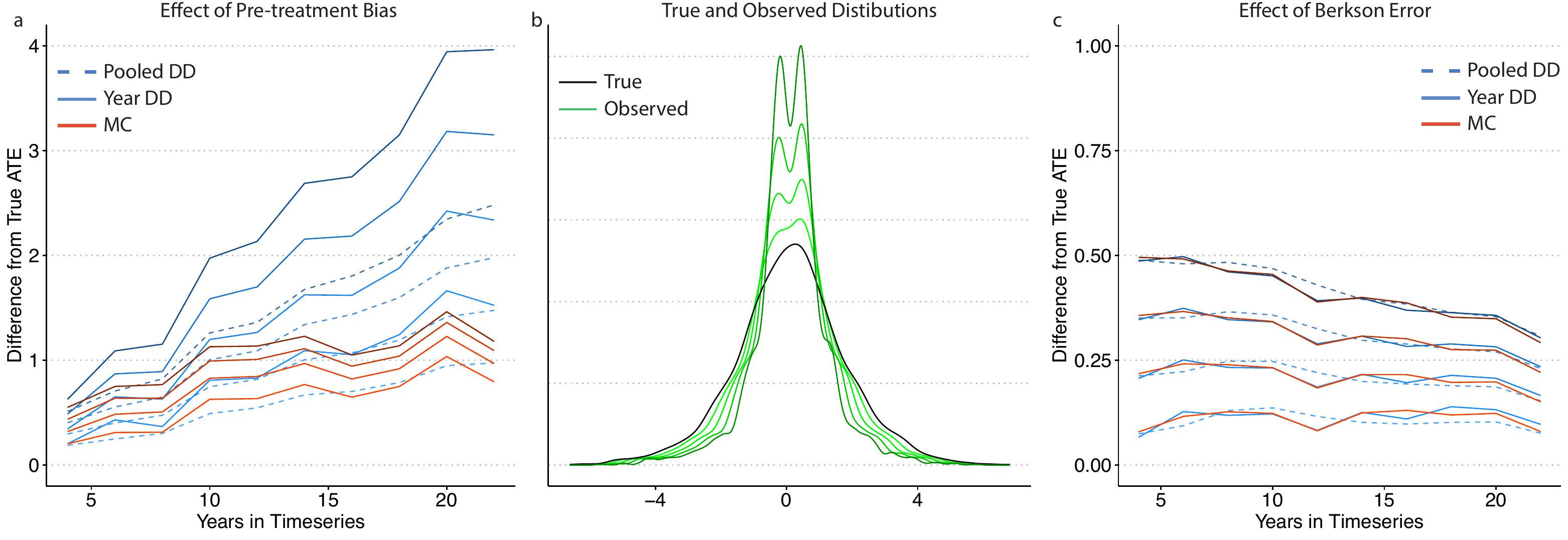}
    \caption{\textbf{Matrix completion (MC) has smaller prediction error than difference-in-differences (DD) when pre-trend bias is present, and both estimators are equally attenuated under Berkson-type measurement error}. \textbf{a}, We simulate a setting where a treatment has a true positive effect equal to 1, but where units are trending differently absent treatment and can have unit-specific trends that are correlated with treatment. We test this situation across multiple time periods (4 to 22 years, where treatment can start in T/2 + 1) and varying growth rates of upward trending units being selected into the treated group (.1 to .25). We then compare DD and MC causal estimates and find that MC is less affected by time-trending unobservables. \textbf{b}, We create a sample of 'observed' data that has a smaller variance than the true distribution, representing a common challenge in data imputation. \textbf{c}, Using our 'observed' sample data, we again simulate a setting that has a positive effect equal to 1 but no pre-trend bias. Under this Berkson-type error, DD and MC are similarly attenuated, underscoring the importance of bias reduction across the outcome distribution in the CNN training process.}
    \label{fig:DD_MC_plot}
\end{figure}

\begin{figure}[H]
\renewcommand{\figurename}{Extended Data Figure}
    \centering
    \includegraphics[width=\linewidth]{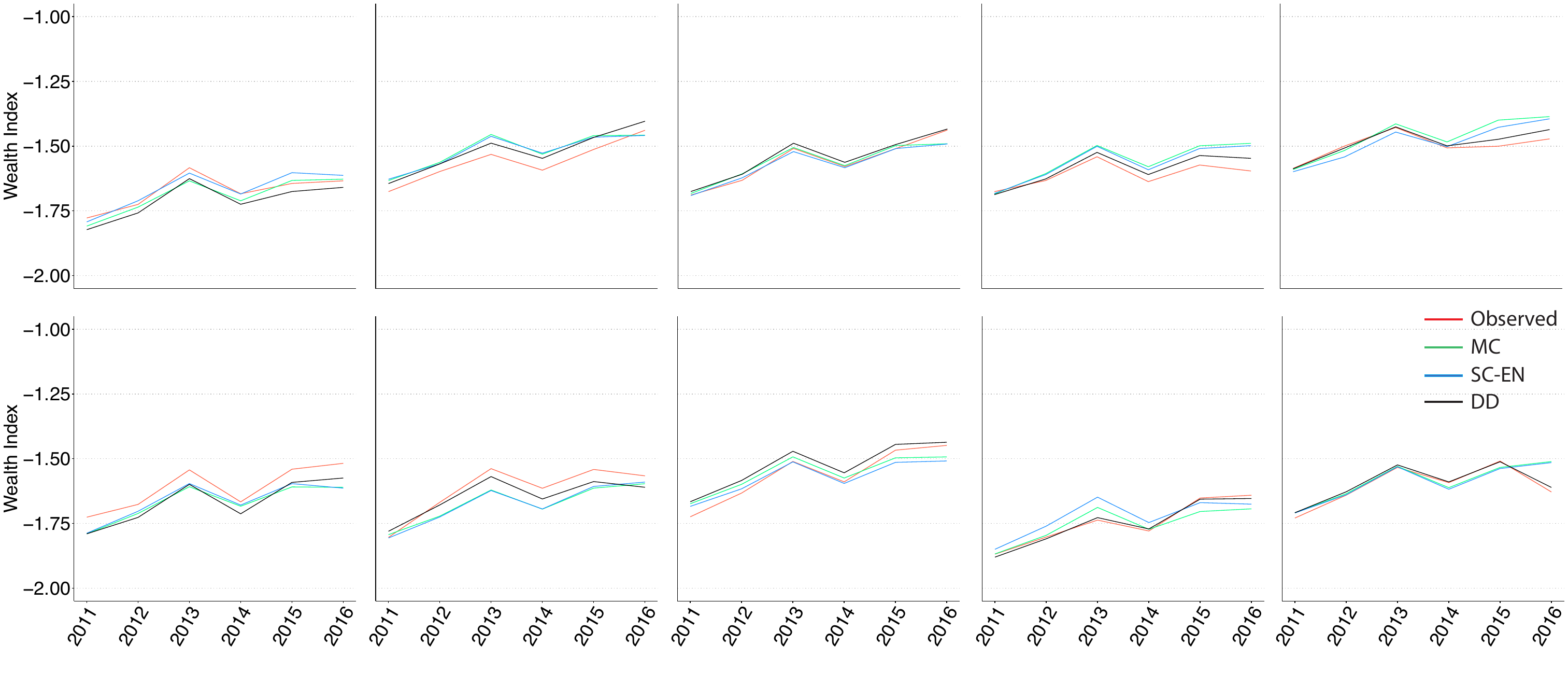}
    \caption{\textbf{Cross validation shows that matrix completion and synthetic controls with elastic net can estimate the average of observed control values with high precision}. We split the control sample into 10 equally sized, random folds, and use matrix completion (MC), synthetic controls with elastic net (SC-EN) and difference-in-differences (DD) to predict cluster-specific wealth values in the held-out test fold for each year, 2011-2016. Using each fold as a test set one time and in the training set nine times.  The mean average difference across all years are .002, .002, and .002 respectively, signifying that MC and SC-EN can predict the average control value with high accuracy.}
    \label{fig:Cross_validation}
\end{figure}

\begin{figure}[H]
\renewcommand{\figurename}{Extended Data Figure}
    \centering
    \includegraphics[width=10cm]{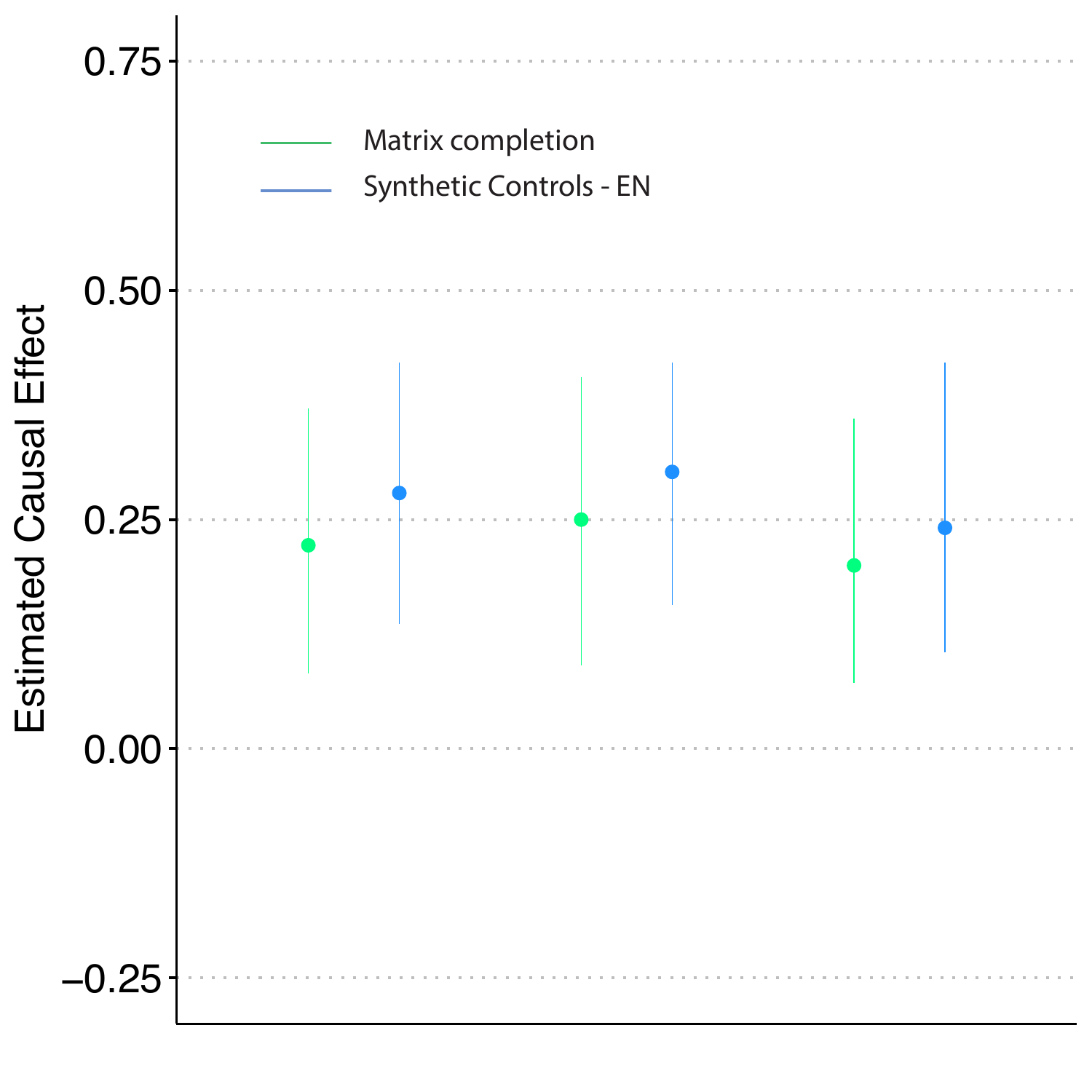}
    \caption{\textbf{Using larger inclusion buffers around grid locations to identify treatment and control samples returns similar causal effect estimates}. We use a 2km buffer around the grid lines as our preferred range to identify treatment and control units as a 2km buffer limits false positives when compared to larger ranges. We also run our causal models using 3km and 4km buffers for comparative purposes.  We find a similar but slightly larger effect at 3km, and similar results for 4km as well.  These results suggest our treatment effect interpretation is robust to buffer range selection.}
    \label{fig:3km_4km_results}
\end{figure}

\begin{figure}[H]
\renewcommand{\figurename}{Extended Data Figure}
    \centering
    \includegraphics[width=10cm]{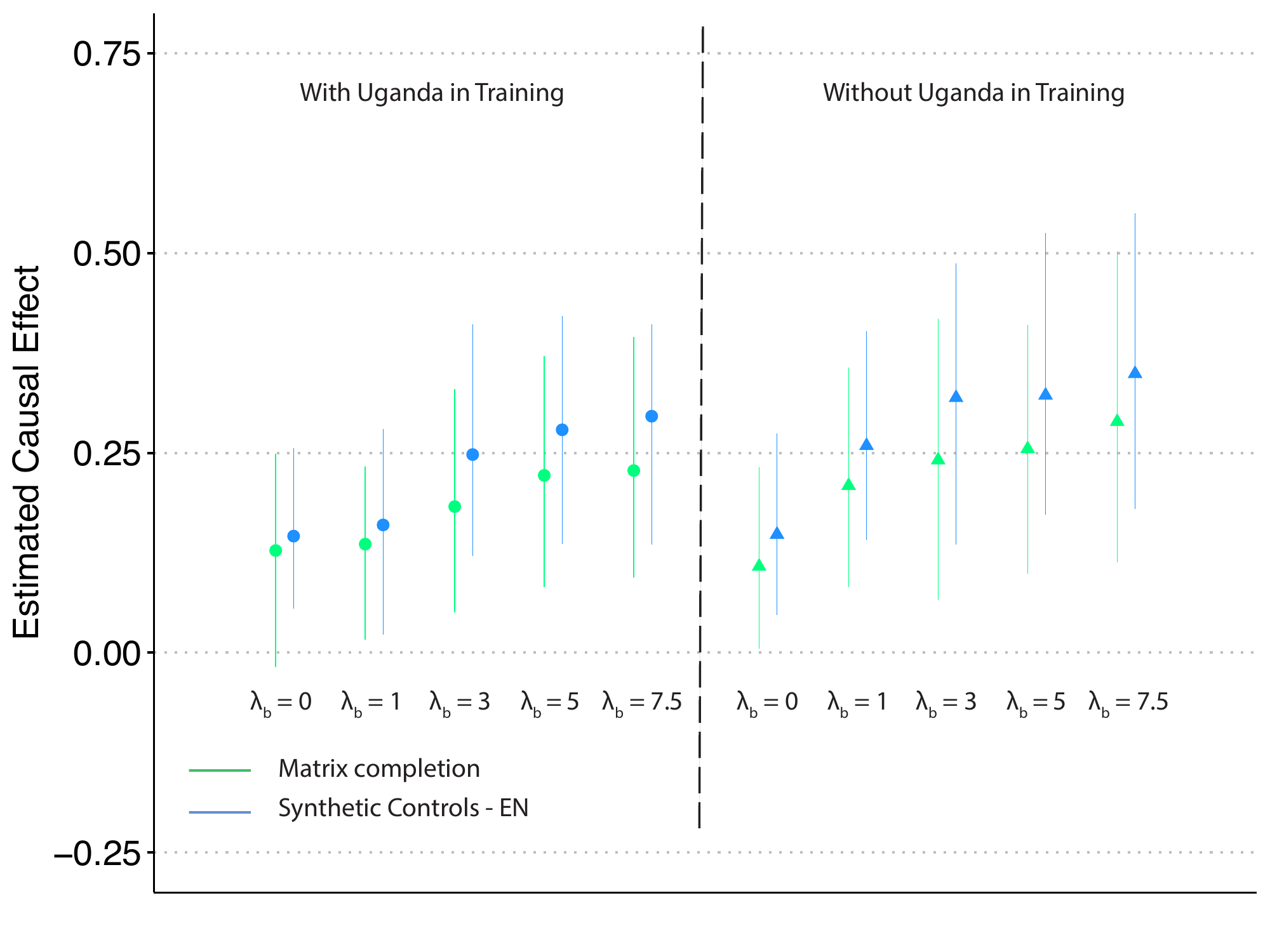}
    \caption{\textbf{The estimated treatment effect of electrification is higher when quintile-specific bias in wealth predictions is penalized more heavily.} Each estimate represents a separate estimate from MC or SC-EN using output from a model with varying quintile-specific bias, from $\lambda_b = 0$ up to $\lambda_b = 7.5$, as in ED Fig \ref{fig:Bias_correction}. 
    }
    \label{fig:Comparitive_treatment_effects}
\end{figure}

\end{document}